\documentclass{webofc}
\usepackage[varg]{txfonts}   
\wocname{ RoPACS conference}
\woctitle{Hot Planets and Cool Stars}
\begin{document}
\title{ Space based microlensing planet searches}
%
%

\author{ Jean-Philippe Beaulieu\inst{1,2}\fnsep\thanks{\email{beaulieu@iap.fr}} \and
              Patrick Tisserand\inst{3}\fnsep\thanks{\email{tisserand@mso.anu.edu.au}} \and
              Virginie Batista\inst{1,4}\fnsep\thanks{\email{virginie@osu.edu}} 
}

\institute{        Institut d’Astrophysique de Paris, Universit\'e Pierre et Marie Curie, UMR7095 UPMC–CNRS 98 bis boulevard
Arago, 75014 Paris, France.
\and   Department of Physics and Astronomy, University College London, London WC1E 6BT, UK
\and   Research School of Astronomy and Astrophysics, Australian National University, Cotter Rd, Weston Creek, ACT 2611, Australia
\and   Department of Astronomy, Ohio State University, Columbus OH 43210, USA
          }

\abstract{%
The discovery of extra-solar planets is arguably the most exciting development in astrophysics during the past 15 years,
 rivalled only by the detection of dark energy. Two projects unite the communities of exoplanet scientists and cosmologists: 
the proposed ESA M class mission EUCLID and the large space mission WFIRST, top ranked by the Astronomy 2010 Decadal 
Survey report. The later states that: ``Space-based microlensing is the optimal approach to providing a true statistical 
census of planetary systems in the Galaxy, over a range of likely semi-major axes”. They also add: “This census, combined 
with that made by the Kepler mission, will determine how common Earth-like planets are over a wide range of orbital parameters''.
We will present a status report of the results obtained by microlensing on exoplanets and the new objectives of
the next generation of ground based wide field imager networks. We will finally discuss the fantastic prospect offered 
by space based microlensing at the horizon 2020-2025.
}
\maketitle
\section{Microlensing planet hunting : where are we in late 2012 ?}
\label{intro}
The number of exoplanets discovered during the last fifteen years is now above 850 (and about 2300 candidates from Kepler), with a sharp increase in the last years. These discoveries have already challenged and revolutionized our theories of planet formation and dynamical evolution. Several methods have been used to find exoplanets: radial velocity, stellar transits, direct imaging, pulsar timing, transit timing, astrometry and gravitational microlensing. Gravitational microlensing is based on Einstein’s theory of general relativity (Gould \& Loeb, 1992): a massive object (the lens) will bend the light of a bright background object (the source). This can generate multiple distorted and magnified images of the source. At the scale of our galaxy, when the lens is a star and the background source is a star located in the Galactic Bulge, these images can not be resolved yet, but the brightness of the source is amplified.
The source’s apparent brightness varies as the alignment changes due to relative proper motion of the source with respect to the lens. Thus, a microlensing event is a transient phenomenon with a typical time scale of $\sim 20 \sqrt{M/M_\odot}$ days and its brightness is monitored to study the event. If the lens is not a single star (i.e. a binary star or star with a planet), the companion will distort the gravitational lens creating regions of enhanced magnification (caustics), which introduce anomalies in the light curve, lasting for about a day for a Jupiter mass and less than two hours for an Earth mass planet.

Microlensing is a rare phenomenon (towards the Galactic Bulge the optical depth to microlensing is  10$^{–6}$). 
Therefore a two-step approach has been adopted since the 1990s. First, wide field imagers are monitoring a very large number of stars in order to detect real time on going microlensing events and alert them publicly (OGLE and MOA collaborations). The second step is to have a network of telescopes (mainly PLANET, $\mu$FUN, RoboNET, Mindstep) doing a follow up of a selected sample of the events to increase the monitoring cadence and then sensitivity to exoplanets. From a network of few telescopes in 2002, we now have up to 50 telescopes available on alert, ranging from robotic 2m telescopes to amateur telescopes in a backyard. In some cases, more than 20 telescopes have been collecting scientifically useful data on a given microlensing event (Batista et al., 2009). Up to now, 19 exoplanets have been published with this method.

This includes cold Neptunes (Gould et al., 2006, Sumi et al. 2010), cold super Earths (Muraki et al. 2011, Bennett et al. 2008, Kubas et al. 2012), Saturns (Bachelet et al.  2012a), Saturn in the Bulge (Janczak et al. 2010), and multiple planet systems (Gaudi et al. 2008, Han et al. 2013). We also detected Brown dwarfs orbiting M dwarfs (Bachelet et al., 2012b) and 3 massive Jupiters orbiting M dwarfs (Dong et al. 2009, Batista et al. 2011, Street et al., 2013) that are not predicted by the core accretion theory (Ida \& Lin 2005, Alibert et al. 2005). On the other hand, gravitational instability can form large planets around M dwarfs (Boss 2006), but typically farther out. Planets formed by such mechanism would have to migrate significantly.

Although the number of microlensing planets is relatively modest compared with that discovered by the radial velocity method and by Kepler, this technique probes a part of the parameter space (host separation vs. planet mass) which is not accessible currently to other methods. The radial velocity and transit method favour the detection of close in and therefore hot planets with a current bias to large/massive planets. More recently it extended to hot super Earths such as GJ1214b, Kepler 10b, hot Earths Kepler 20e and 20f and the first large sample of terrestrial-sized exoplanets (Buchhave et al. 2012). Moreover, within the 2000 planetary systems discovered by Kepler, 50 exoplanets are in the stellar habitable zone. Microlensing complements these detections because it is most sensitive to planets beyond the distance where water ice forms (the snow line), and to masses down to the Earth. Gould et al. (2010) have made the first measurement of the frequency of icy and gaseous planets beyond the snow line, and have shown that this is about 7 times higher than closer-in systems probed by the Doppler method (Cumming et al. 2010).  This comparison provides strong evidence that most giant planets do not migrate inwards very far. Howard et al. (2011) has presented the first abundances of planets orbiting solar like stars within 0.25 AU using Kepler, while Mayor et al. (2011) have measured the abundance of Neptune and super Earth using radial velocities. These studies show that 17-30 \% of solar like stars have planets on short orbits. Cassan et al. (2012) finds that $17_{-9}^{+6}$  \% of stars host Jupiter-mass planets (0.3-10 $M_J$). Cool Neptunes (10-30 $M_\oplus$) and super-Earths (5-10 $M_\oplus$), however, are even more common: their respective abundances per star are $52_{-29}^{+22}$ \% and $62_{-37}^{+35}$ \%. Planets around stars are the rule, rather than the exception. 

The derived mass function is   $\rm d N /(d \log a ~ d \log M) =  10^{-0.62 \pm 0.22} ~ (M/M_{sat})^{-0.73 \pm 0.17}$ (where N is the average number of planets per star, a the semi-major axis and M the planet mass, and $M_{sat}=95M_\oplus$), appears steeper than results from the Doppler technique and abundances slightly larger. Differences can arise because the Doppler technique focuses mostly on solar-like stars, whereas microlensing a priori probes all types of host stars. Moreover, microlensing planets are located further away from their stars (closer to the locus of formation) and are cooler than Doppler planets (that have been almost certainly affected by migration). Bonfils et al., (2013) just released their statistics on the M dwarf sample monitored by HARPS on orbits between 1-100 days and concluded a high abundance of  super Earth, in agreement with the microlensing result (although probing inner orbits).  

Microlensing is roughly uniformly sensitive to planets orbiting all types of stars, as well as white dwarfs, neutron stars, and black holes, while other methods are most sensitive to FGK dwarfs and are now extending to M dwarfs. It is therefore an independent and complementary detection method for aiding a comprehensive understanding of the planet formation process. It is currently capable of detecting cool planets of super-Earth mass from the ground and, with a network of wide field telescopes strategically located around the world, could detect planets with mass as low as the Earth. 

Exoplanets probed by microlensing are orbiting stars much further away than those probed with other methods, which are sensitive to planets orbiting stars in the Sun neighborhood. They provide an interesting comparison sample with nearby exoplanets, and allow us to study the extra-solar population throughout the Galaxy, in the Disk but also in the  galactic Bulge (Janczak et al. 2010). In particular, the host stars with exoplanets appear to have higher metallicity (Fischer and Valenti, 2005). Since the metallicity is on average higher as one goes towards the Galactic Center, the abundance of exoplanets may well be somewhat higher in microlensing surveys. Ground-based microlensing mostly probes exoplanets outside the snow line, where the favored core-accretion theory of planet formation predicts a larger number of low-mass exoplanets (Ida \& Lin, 2005, Alibert el al. 2005). 

Since microlensing can instantaneously detect planets without waiting for a full orbital period, it is immediately sensitive to planets with very long periods. Although the probability of detecting a planet decreases for planets with separations larger than the Einstein ring radius, it does not drop to zero. As a consequence, planets on very wide orbits, and free-floating planets (ejected during the formation process or formed as such) are detectable by microlensing. A significant population of free-floating planets is a generic prediction of most planet formation models, particular those that invoke strong dynamical interactions to explain the observed eccentricity distribution of planets (Goldreich et al., 2004, Juric \& Tremaine, 2008, Ford \&  Rasio, 2006).  An important population of free-floating Jupiters has also been unveiled using the microlensing technique (Sumi et al. 2011) and is raising questions about the importance of dynamical interactions after the formation of the planets (Morbidelli et al., 2012). In the coming years it will be important to check their free-floating nature (not bound to a star), to perform mass measurements when possible (thanks to terrestrial parallax, Gould et al. 2010), to measure their abundance and to detect lower mass ones. Recently, the first direct detection of a free-floating Jupiter-mass planet was revealed (Delorme et al., 2012). 

Most planets in our Solar System are surrounded by satellites or moons, some of them being also part of a binary system (Pluto and Charon). 
The population of multiple-planetary systems containing many planets is increasing (e.g. Kepler 11, HD10180, Gliese 581).  Although exomoons 
have not been published yet, various dedicated methods have been proposed for their detection, such as transit light curve, transit timing, direct 
imaging, microlensing (Bennett \& Rhie 2002, Liebig \& Wambsganss, 2010) and Doppler spectroscopy. 
Kepler's high precision has opened the possibility of detecting exomoons (Kipping et al., 2009). Steffen J.H., et al., (2013) 
is presenting the most up to date transit timing work by the Kepler team, while an independent survey using public Kepler data is underway, the Hunt for Exomoons with Kepler (Kipping et al. 2012). Several analyses for exomoon detection have been published but none of them revealed any evidence for moon signature (Kipping et al. 2013, Nesvorny et al. 2012, Montalto et al. 2012). Recent work on MOA 2011-BLG-262 has shown that, with microlensing technics, it is already possible to detect exomoons of the mass of Mars orbiting free floating planets and around distant planets from their star. After this pioneer detection (to be published most likely spring 2013), as part of EARTH-HUNTER, we will do a systematic search for such systems. 

The statistics provided by microlensing, combined with those from other methods, will thus enable a critical test of planetary formation models (Ida \& Lin 2005, Mordasini et al., 2009). Microlensing can provide a census of cold planets that matches in sensitivity and extends the parameter space of other large surveys conducted using transits and radial velocities. 

\section{The network of wide field imagers network era (2012-2020)} 

The existing structure of a network of telescopes controlled by PLANET, $\mu$FUN, OGLE, MOA, RoboNET and MINDSTEP was netting 5-7 planets per year in the period 2007-2011. In 2012, 22 planets have been discovered. Contrary to the previous years, the wide field imager contribution from OGLE-IV, MOA-II and WISE is entirely dominant over the contribution of the fleet of the follow up telescopes for most of the planets. We are very clearly at a turning point, as it has been anticipated years ago (Gaudi et al., 2009). Moreover, other wide field imagers will join the worldwide effort from 2013. The first ones will be the new Harlingten 1.3m telescope at the Greenhill observatory (Bies Die Tier, Tasmania) and the 1.3m SkyMapper telescope at Siding Spring Observatory. They will be operated as part of the EARTH-HUNTER collaboration.  In the period 2014-2016, the 3 telescopes from the KMTNet will also come online at CTIO, SAAO and Siding Spring. Each node will be a 1.6m telescope equipped with 4 square degrees camera. Moreover, there are also plans to install a wide field imager in Namibia. Meanwhile, the LCOGT telescopes is deploying 1m and 40 cm telescopes around the world. They could be an interesting addition to the wide field imager network by performing simultaneous
observations from different latitude/longitude to allow for detection of terrestrial parallax that will help mass measurements (Gould \& Yee  2013). 
In addition to the wide field imager photometry, there is a need to perform the coordination of targeted observations with Adaptive Optics systems on KECK, SUBARU, VLT and HST in order to add constrains on the brightness of the lens to better determine the parameters of the system (Donatowicz, et al., 2008).
Observations few years apart will also permit to measure the direction and amplitude of the proper motion lens-source, allowing to nail down
the parameters accurately.

With this worldwide effort, at the horizon 2018, the following objectives will be reached :  \\
 \noindent 1) Measure the frequency of Earth-mass planets beyond the snow line \\
 \noindent 2) Measure the frequency of free-floating (i.e., ejected) gaseous planets \\
 \noindent 3) Measure the frequency of giant planets beyond the snow line as a function of planet-host mass
and separation \\
 \noindent  4) First constrains the frequency of exomoons via microlensing.


\section{The ultimate planet hunting machines, EUCLID and WFIRST}

At the horizon 2020+, a wide field imager in space will obtain a comprehensive census from free-floating small mass telluric planets to frozen Mars and habitable Earth orbiting solar like stars. The concept was initiated with a dedicated mission (Bennett and Rhie 2002, Bennett et al., 2012) called the Microlensing Planet Finder (MPF), which has been proposed to NASA's Discovery program but not selected.  
The objective is to be able to monitor turn off stars in the galactic bulge as microlensing sources. Bennett and Rhie (1996) had shown that the detectability of exoplanets via microlensing depends strongly on the size of the source star. As an example, the $\sim 5.5\rm  M_\oplus$ super Earth detected by Beaulieu et al. (2006) with a bulge giant as a source star was close to the limit of detection. Lower mass planets are detectable only with small source star, unresolved from the ground. 
Given the extinction of the fields and the temperature of the source stars, the monitoring is better done in the IR. Moreover, about $\sim 2 \rm deg^2$ towards the galactic Bulge have an optical depth to microlensing 3 to 4 times higher than the average fields actually monitored in the optical by ground based surveys. The ideal microlensing planet hunting machine is a 1m class telescope, with a wide field IR imager (about $\sim 0.5 \rm deg^2$ and high angular resolution ($0.1-0.3$arcsec/pix), to observe the galactic bulge with a sampling rate better than 20 min. Although it could be done in the optical, the highest efficiency will be at H band.
Despite the fact that the designs were completely independent, there is a remarkable similarity  between the requirements for missions 
aimed at probing Dark Energy via cosmic shear, baryonic acoustic oscillations and a microlensing planet hunting mission. The requirements
of the designs are stronger for the cosmic shear compared to the microlensing.

\begin{table}
\caption{A short summary of EUCLID and WFIRST. Please note that although the design of EUCLID 
is definite, the design from WFIRST is under development. We discuss the opportunity
of using 2.4m NRO telescope for WFIRST in the text. }
\label{tab-1}       
\begin{tabular}{ll}
\hline
EUCLID & WFIRST, DRM1 and DRM2 \\\hline
ESA M2 mission launch 2020 & NASA mission, launch 2025+? \\
1.2m Korsch telescope & 1.3m off axis, three-mirror anastigmat telescope\\ \hline
Optical, 0.1 arcsec/pix, FOV 0.54 deg$^2$ & $0.92-2.4 \mu$m\\ 
IR bands : Y, J, H, 0.3 arcsec/pix, FOV 0.58 deg$^2$ & IR bands : Z, Y, J, H, K, W bands, 0.19 arcsec/pix,  \\
                                                                                   & FOW 0.375 deg$^2$ (DRM1)  or $\sim 0.585$ deg$^2$ (DRM2) \\ 
\hline
                                     Core science : Dark Energy.   &  Core science :  Dark Energy, exoplanets, \\
                                                                                   &  Galactic science and general observatory.\\
Legacy science : Exoplanets, SNIa, Galactic science   & \\
\hline
 4-6 months microlensing program (2020-2026) & 400-500 days microlensing program but when ?  \\
followed by 6 months after 2027 ?                        & launch date depending on budget after JWST \\
To be decided in 2014-2015                                  & launch. \\
\\ \hline
\end{tabular}
\end{table}

EUCLID is an ESA medium size mission. As core science, it will measure parameters of dark energy using weak gravitational  lensing and baryonic acoustic oscillation, test the general relativity and the Cold Dark Matter paradigm for structure formation. Since its original submission (under the brand DUNE) in 2007 to ESA, a microlensing planet hunting program has been listed as part of the Legacy  science (Beaulieu et al. 2007, 2008, 2010). 

The vision adopted by the Europeans of a joint mission with Dark Energy probes and microlensing has been promoted and adopted by the Astro 2010 Decadal Survey 
when it created and ranked as top priority the WFIRST mission. The report stated : 
{  ``Space-based microlensing is the optimal approach to providing a true statistical census of planetary systems in the Galaxy, over a range of likely 
semi-major axes”. } They also added: {  “This census, combined with that made by the Kepler mission, will determine how common Earth-like planets are over a wide range of orbital parameters''.}
Some characteristics of EUCLID and WFIRST are summarized in  table~1, and we will discuss the microlensing capability of these missions
in the forecoming two sections.

\subsection{EUCLID}

The EUCLID telescope is scheduled for launch in 2020 (Laureijs R. et al. 2011). It consists of a 1.2 m Korsch telescope designed to provide a large field of view, 
equipped with two instruments that will observed simultaneously the same field of view ($\sim$ 0.54 deg$^2$) via a dichroic filter. 
The reflected light is led to the visual instrument (VIS) and the transmitted light from the dichroic feeds the near infrared instrument (NISP) which 
contains a slit-less spectrometer and a three bands photometer (Y, J, H).  The VIS camera is made of 36 CCDs of 0.1 arcsec per pixel and will 
detect light over a wide visible band (R+I+Z). The NISP photometer uses 16 HgCdTe NIR detectors of 0.3 arcsec per pixel.  It should take 6 years 
to complete the main scientific mission, that is a wide survey which covers 15,000 deg$^2$ of the extra-galactic sky down to AB mag 24.5, 
complemented by two 20 deg$^2$ deep fields observed on a monthly basis. The current observing strategy of the main EUCLID survey 
means that for up to two months per year it will point towards the Galactic plane and away from its primary science fields. Some of this 
time is intended to be devoted to other legacy science. 

A baseline for the design of the EUCLID microlensing survey is described by Penny et al. (2013a) with a detailed simulation demonstrating 
the capabilities and the scientific outcomes expected. It uses  a multi-wavelength microlensing simulator, a realistic galactic population 
distribution based on the Besan\c con galactic model (Robin et al., 2003) and current numerical models of the EUCLID PSF. In summary, 
the simulation shows that it will be the first survey able to measure the abundance of exoplanets down to Earth mass for
 host separations from $\sim$1 AU out to the free-floating (unbound) regime. 

The program propose to monitor 3 contiguous fields located toward the galactic bulge (i.e. $\sim$1.6 deg$^2$) with a sampling 
rate of 17 minutes for one month continuously.  The survey would be conducted primarily in the H band as it yields 
to the largest number of planet detections, with Y and J band taken every 12 hours. 
This one month observing sequence will be repeated 4-6 times during the nominal  6 years of the EUCLID mission, and then 
in the extended part of the mission. The exact amount of telescope time allocated to this project will be decided in 2014-2015.
 Penny et al. (2013a) shows that already with a 4 months survey the slope of the mass function will be constraint strongly down to 
small masses. In order to reach the habitable zone around solar like stars overlapping with the results from Kepler, a 10+ months
survey would be needed, mainly in the extended part of the EUCLID mission. It is truly remarkable that a small allocation of time 
on EUCLID will be able already to give strong constraints on planetary formation models.

\subsection{WFIRST}

Two reference designs have been originally considered for WFIRST (Green et al. 2012) to meet the requirements from the decadal survey (table~1). DRM1 is the first design to address all the science goals as a
stand alone mission. DRM2 is non-duplicative of EUCLID and it 
is a cheaper version (1 billion versus 1.6 billions) using 4k chips H4RG instead of the 2k version.
It is a pure imaging mission with a wider field of view of $\sim 0.585 \rm deg^2$. DRM1 has a smaller 
field of view, but two spectroscopic modes using a prism and a grating. The microlensing capability 
of WFIRST has been investigated first by Barry et al., (2011) and a  thorough study is underway by Penny et al. (2013b).
The later will apply exactly the same codes and hypothesis as done by Penny et al. (2013a) but for WFIRST.

NASA has acquired two complete 2.4m telescopes from the National Reconnaissance Organization (NRO) 
of the US Department of Defense. They are wide field versions of the Hubble Space Telescope. Dressler et al., (2012)
has studied the opportunity to use one of these telescopes to meet the Astro 2010 Decadal survey goals, and called it the NEW-WFIRST project. Provided that existing hardware would be used, it is envisioned it would both fit in the
DRM1 cost cap, while  addressing all the objectives from the Decadal survey in a shorter time scale compared to 
the DRM1 implementation. Nevertheless, although it is promising, more studies are needed to confirm that indeed
it is possible.
For microlensing, the planet catch with NEW-WFIRST is estimated to be $\sim 1.6$ times larger than DRM1 design, 
with a higher efficiency in particular towards the low mass planets ($\sim $3 times more Mars mass planets). These
estimates (not yet a full simulation) have been done with a survey of the same duration, with a IR focal plane of the 
same size, but smaller pixels for the NEW-WFIRST version.

Even if the European time scales are known to be rather slow, WFIRST will most likely be launched when 
EUCLID will be finishing his 6 years core survey. Indeed, it is only after JWST launch that WFIRST will be implemented (~2017).
As a consequence, there is time to estimate the performance of the NEW-WFIRST project with respect to the previous
small brother version called DRM1.

\section{Conclusion}
Microlensing is at its golden age. It has a specific niche to address scientific questions that can only be answered by this approach. 
The different consortium involved in the ground base studies have evolved from competitive teams, to a global worldwide super consortium
that goes beyond the standard approach. This cooperation/competition has been the key of the success and should be taken as a model
by the other planet hunting techniques.  The road is clear, with statistics of frozen Earth (among the objectives) to be obtained in the coming five years
from the ground thanks to OGLE-IV, MOA-II, KMTNet, EARTH-HUNTER and RoboNET. Finally, a census on exoplanets down to the mass of Mars,
including free floating planets will be done by EUCLID and WFIRST.

\end{document}